\documentclass[aps,prev,twocolumn,preprintnumbers,floatfix,nofootinbib]{openjournal}
\pdfoutput=1

\usepackage{mathrsfs}
\usepackage{amsmath, amsthm, amssymb}
\usepackage{color}

\newcommand{\D}[1]{\ensuremath{\text{d}#1}}

\newcommand{\Exp}[1]{\ensuremath{\textrm{e}^{#1}}}
\newcommand{\E}[1]{\ensuremath{\cdot10^{#1}}}

\newcommand{\Inf}{\ensuremath{\varphi}}
\newcommand{\Ax}{\ensuremath{\phi}}
\newcommand{\Del}[1]{\ensuremath{\partial_{#1}}}
\newcommand{\DEL}[1]{\ensuremath{\partial^{#1}}}
\newcommand{\T}{\ensuremath{\tau}}
\renewcommand{\vec}[1]{\ensuremath{\mathbf{#1}}}
\newcommand{\R}{\ensuremath{\mathcal{R}}}
\renewcommand{\P}{\ensuremath{\mathcal{P}}}
\renewcommand{\L}{\ensuremath{\mathcal{L}}}

\newcommand{\arsinh}{\operatorname{arsinh}}

\renewcommand{\Im}{\operatorname{Im}}
\renewcommand{\Re}{\operatorname{Re}}
\newcommand{\be}{\begin{equation}}
\newcommand{\ee}{\end{equation}}
\newcommand{\bea}{\begin{eqnarray}}
\newcommand{\eea}{\end{eqnarray}}

\begin{document}

\title{Squeezing the Axion}

\author{Jondalar L. J. Ku{\ss}$^a$}
\email{j.kuss@stud.uni-goettingen.de}
\author{David J. E. Marsh$^b$}
\email{david.j.marsh@kcl.ac.uk}

\vspace{1cm}
\affiliation{$^a$Institut f\"{u}r Astrophysik, Georg-August Universit\"{a}t, Friedrich-Hund-Platz 1, D-37077 G\"{o}ttingen, Germany}
\affiliation{$^b$Theoretical Particle Physics and Cosmology Group, Department of Physics, King's College London, Strand, London WC2R 2LS, U.K.}

\begin{abstract}

We apply the squeezed state formalism to scalar field dark matter (e.g. axion) perturbations generated during inflation. As for the inflationary perturbations, the scalar field state becomes highly squeezed as modes exit the horizon. For as long as $H>m_\Ax$ (with $H$ the Hubble rate and $m_\Ax$ the scalar mass) the scalar field field does not interact during reheating, and we follow its evolution exactly as modes re-enter the horizon. We find that the quantum state remains squeezed after horizon re-entry during the hot big bang. This demonstrates a fact well-known in the theory of inflation: cosmological observables for scalar dark matter are accurately modelled by a classical stochastic field with a fixed phase. Our calculation covers all modes smaller than the present-day cosmic de Broglie wavelength. Larger scale modes mix gravitationally with the environment when $H<m_\Ax$, and are thus expected to decohere. \\

Preprint: KCL-PH-TH/2021-37

\end{abstract}

\maketitle
\section{Introduction}
\label{sec:introduction}

The history of the Universe in the standard cosmological model is believed to proceed from an initial period of inflation~\citep{1981PhRvD..23..347G,1982PhLB..108..389L,1982PhRvL..48.1220A}, during which the Universe is in a quasi-de Sitter state with weakly broken scale invariance. During this epoch, the Friedmann-Robertson-Walker (FRW) scale factor, $a$, grows exponentially fast with respect to cosmic time, solving the horizon and flatness problems, as required by observations of the cosmic microwave background (CMB) anisotropies~\citep{Akrami:2018vks}. Inflation also provides initial conditions for the small perturbations in the CMB, which arise from quantum vacuum fluctuations of the inflaton field, $\Inf$. However, the CMB sky is classical, and it is extremely challenging to look for cosmic signatures of our quantum past.

The explanation for this classicality is found within the inflation model itself. Quantum fluctuations are ``squeezed'' as they exit the horizon~\citep{Ferreira,Polarski:1995jg}. In the squeezed state, the mode functions can be treated as entirely real, with vanishing commutator. In fact, at the level of Gaussianity, all statistics can be matched exactly by classical stochastic fields, regardless of the level of squeezing, as long as the occupation number is large, i.e. $n\lambda^3\gg 1$ where $n$ is the number density and $\lambda$ is the wavelength \citep[see e.g. ][]{sakurai,Martin:2015qta}. Inflation creates the large particle number from the Bunch-Davies initial state (``pulling particles out of the vacuum''), and the act of squeezing is the dominance of the growing mode of perturbations over the decaying mode, which has the effect of fixing the phase of the classical stochastic model. 

We note that the squeezing picture is somewhat simplistic: in the following we follow no interactions, and the Gaussian states are exactly equivalent to a classical model. However the intuition imparted  from squeezing, that off diagonal terms in the density matrix become vanishingly small and can be neglected, is essentially born out in the open effective field theory/Linblad equation description of \cite{Burgess:2014eoa}.

Once all modes of interest have left the horizon during inflation, a further classicalization occurs. The inflaton field decays and ``reheats'' the other degrees of freedom, producing a thermal bath~\citep{Kofman:1997yn}. All particles, photons, baryons, cold dark matter (CDM), neutrinos, are then created out of this thermal bath. For the most part, the memory of the initial quantum state is carried in the classical stochastic correlations of density perturbations in these components \citep[although see e.g.][for discussions about accessing the quantum nature of inflation using cosmological observables]{Maldacena:2015bha,Lim:2014uea,Martin:2015qta}.

An interesting possible exception to this story of quantum past and classical present is in the case of scalar field dark matter (DM) such as the axion~\citep{Peccei:1977ur,weinberg1978,wilczek1978} and axion-like particles \citep[see][for a review]{Marsh:2015xka}. In the standard treatment of this model, the majority of the DM is created by non-thermal processes~\citep{1983PhLB..120..133A,1983PhLB..120..137D,1983PhLB..120..127P}. Crucially, the DM production requires no direct interactions with the thermal bath, and proceeds only from motion of the scalar field from a displaced initial state to the vacuum. It is described, at least initially, by a pure state.

The evolution of the classical mean field is generally thought to capture all of the relevant physics. The classical field exhibits a de Broglie wavelength, which manifests as a Jeans scale suppressing cosmic structure formation on small scales~\citep{khlopov_scalar}. The classical field undergoes Bose-Einstein condensation, and forms axion stars, or solitons~\citep{2014NatPh..10..496S,Levkov:2018kau}. In regions of coherent flows, such as in cosmic filaments, the classical field also displays interference patterns~\citep{2014NatPh..10..496S,Mocz:2019pyf}. However, it has been an ongoing topic of research in the field of axion DM to ask whether this classical mean field is in fact the complete and correct description, or whether there are truly quantum and beyond mean field effects, i.e. quantum correlations, that must be accounted for~\citep[e.g][]{2009PhRvL.103k1301S,2015PhRvD..92j3513G,Berges:2014xea,Sikivie:2016enz,Hertzberg:2016tal,Dvali:2017ruz,Lentz:2018fqb,Lentz:2019xcr}.

In the following, we attempt to at least partially answer this question in a simple setting. We consider the case where the Peccei-Quinnn symmetry is broken during inflation. In this case, the axion field, $\Ax$, is linear for most of cosmic history, vastly simplifying the computation. In this case, the axion field is smoothed during inflation, giving rise to a background homogeneous condensate, $\Ax_0(t)$. This homogeneous field provides the axion DM relic density from vacuum realigment. Furthermore, such a homogeneous field is described exactly by mean field theory, with no exchange correlation~\citep{Lentz:2018fqb,Lentz:2019xcr}. 

The correlations of perturbations in the field are split into two modes: the adiabatic mode, and the isocurvature mode. The adiabatic axion fluctuations have asymptotically zero fluctuation in modes with $k>0$~\citep{Hlozek:2014lca}, while the $k=0$ mode is described exactly by the classical mean field. All DM structure in this mode is inherited from the the source terms of the thermal bath. Therefore, because these modes are created thermally, like those of the standard cosmological model they will have lost memory of their quantum past. 

We thus look to the isocurvature mode for the axion quantum state. Axion isocurvature fluctuations arise during inflation from vacuum fluctuations of the axion field~\citep{Axenides:1983hj}. Since the axion is so much lighter than the inflaton (typically $m_{\rm \Inf}\lesssim 10^{14}\text{ GeV}$ while $m_\Ax\ll 1 \text{ eV}$), the axion behaves at all times as a spectator field~\citep{Gordon:2000hv} \citep[although see e.g.][]{Guth:2018hsa,Graham:2018jyp,Marsh:2019bjr}. The isocurvature modes leave the horizon just like the inflaton modes and, as we show below, become squeezed. The standard treatment of axion isocurvature perturbations \citep[see e.g.][]{2009ApJS..180..330K,hertzberg2008,Hlozek:2017zzf} treats them like the inflaton and the adiabatic mode, with a quantum past, but a classical stochastic model of the structure formation phase during and after the hot big bang. 

The fundamental difference between the axion and the inflaton (or indeed a more exotic model such as the curvaton~\citep{2002PhLB..524....5L} is that, by necessity if it is to be DM, the axion field is \emph{stable}, and the vacuum realigment relic density is produced without need for any thermal interaction. During the hot big bang phase, modes that were super horizon after inflation re-enter the horizon. The evolution of the homogeneous background field splits the modes into two categories. The background field begins to oscillate when $H\approx m_\Ax$, and it is at this time that the scalar field DM is produced. 

Modes that enter the horizon prior to axion DM production, never interact or mix gravitationally with any external fields (in linear theory). Thus they cannot decohere, and ``unsqueezing'' of the initial pure state could occur. Large scale modes re-enter the horizon when $H<m_\Ax$. The oscillation of the background field serves to ``swicth on'' the interaction with the gravitational potential, and consequent mixing with the thermal bath. This interaction is expected to decohere the large scale modes~\citep{Joos:1984uk} \citep[see also][for decoherence inside galactic halos]{Allali:2020ttz}. The smaller scale modes (which are smaller than the cosmic de Broglie wavelength/Jeans scale), however, might retain memory of their quantum initial state. It is thus in axion isocurvature fluctuations that enter the horizon prior to axion particle production where the initial quantum state could be encoded in the evolution in the form of the phase, which is normally neglected.

In the present work we model this possibility and, in linear perturbation theory, evolve the axion field from the initial state during inflation, through reheating, and horizon re-entry during the hot big bang phase right up until the moment of axion DM production. We are thus able to compute the squeeze parameter of the axion field, including any possible unsqueezing of sub-de Broglie modes during horizon re-entry. 

This paper is organised as follows. In Section~\ref{sec:model} we introduce the basics of our cosmological model, in Section~\ref{sec:squeezing} we present the squeezing model of inflationary fluctuations, and in Section~\ref{sec:results} we compute the squeeze parameter of the axion isocurvature mode and its evolution through horizon reentry. We conclude and discuss our results in Section~\ref{sec:Discussion}.

\section{Model}
\label{sec:model}
\begin{figure}[t]
  \centering
  \input{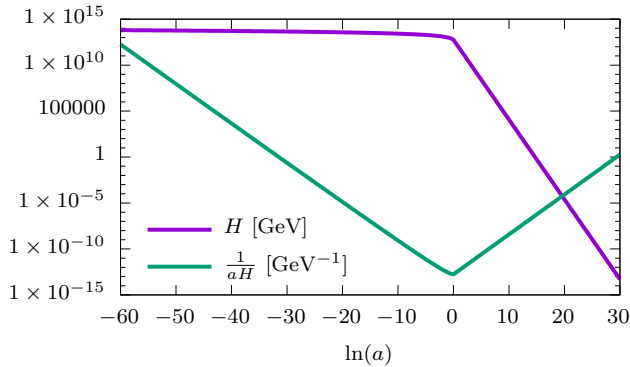}
  \caption{Evolution of Hubble parameter $H$ and comoving Horizon $1/aH$ in the background model.}
  \label{fig:background}
\end{figure}

We employ a simple toy model for the background evolution consisting of single field slow-roll inflation with a quadratic potential $V(\Inf) = \frac{1}{2}m_\Inf^2\Inf^2$. This choice is made purely for simplicity of the analytic background solution. Our conclusions concerning the axion quantum state are not strongly affected by this choice as long as 1) Peccei-Quinn symmetry is broken during inflation and not restored afterwards and 2) the axion remains a spectator field. Inflation is followed by instantaneous reheating leading to a radiation dominated epoch (i.e. the hot big bang). The FRW scale factor, $a(t)$, is then given by
\begin{equation}
    a(t) = 
    \begin{cases}
    a_i \Exp{2\pi G (\Inf_i^2 - \Inf^2(t))} & \text{inflation} \\
    \sqrt{\frac{2m_\Inf}{\sqrt{3}}} T^{\frac{1}{2}} & \text{radiation}
    \end{cases},
    \label{eq:a-of-t}
\end{equation}
with $a_i = \Exp{2\pi G (\Inf_\text{reheat}^2-\Inf_i^2)}$ and $T = t-t_\text{reheat}-\frac{\sqrt{3}}{2m_\Inf}$. Reheating occurs for $\Inf_{\text{reheat}} = \frac{1}{\sqrt{4\pi G}}$, the scale factor is normalised to unity here, $a_\text{reheat}=1$.  The mass of the inflaton is taken to be $m_\Inf = 10^{13}\,\text{GeV} \approx 10^{-6}M_P$~\citep{peter05} (where $M_P=1/\sqrt{8\pi G}$), the initial field value is chosen as $\Inf_i=\tfrac{4}{\sqrt{G}}$ yielding approximately one hundred e-folds of inflation. 
For the numerical treatment, it is convenient to use the natural logarithm $\ln(a)$ of the scale factor as the time variable. We can then express the Hubble parameter as
\begin{equation}
    H =
    \begin{cases}
    m_\Inf \sqrt{\frac{1}{3}-\frac{2}{3}\ln(a)} & a \leq 1 \\
    \frac{m_\Inf}{\sqrt{3}}\Exp{-2\ln(a)} & a \geq 1
    \end{cases}.
\end{equation}
The evolution of $H$ and the comoving horizon is shown in Fig.~\ref{fig:background}.

In this background model, the axion does not play any role: it is assumed to contribute only negligibly to the total energy density, which hence is dominated by the inflaton or by radiation respectively. 
The background evolution of the homogeneous axion field is given by:
\begin{equation}
    \ddot{\Ax} + 3H\dot{\Ax}+m_\Ax^2\Ax=0,
    \label{eq:eom-axion}
\end{equation}
where we have assumed that the initial displacement of the axion field is less than the Peccei-Quinn symmetry breaking scale, $\Ax<f_a$, such that the axion potential is approximately quadratic. Again, this choice is motivated purely by simplicity and our conclusions concerning the quantum state would not be strongly affected by using the full axion potential. We approximate the axion mass as constant, as for a generic axion-like particle. A temperature dependent mass, as for the QCD axion, would also not change the character of our conclusions, which apply in the limit $H\ll m_\Ax$. Within these approximations, the axion behaves like a damped oscillator and is essentially frozen at its initial value until the Hubble parameter becomes comparable to the axion mass, which we assume happens deep in the radiation dominated era. 

For studying the perturbations we follow \cite{Gordon:2000hv} in separating curvature and isocurvature perturbations. They consider the space spanned by two scalar fields $\chi_1$ and $\chi_2$.
By a rotation in this field space, a perturbation is split into a component along the the trajectory of the background field's evolution and one component perpendicular to it. It turns out that the former, $\delta\sigma = \cos(\theta)\delta\chi_1 + \sin(\theta)\delta\chi_2$ ($\theta$ being the angle between $\chi_1$ and the background trajectory), is associated with curvature perturbations, $\R$ and the latter, $\delta s = -\sin(\theta)\delta\chi_1 + \cos(\theta)\delta\chi_2$, with fluctuations in entropy, $S = H \left( \frac{\delta p}{\dot{p}} - \frac{\delta\rho}{\dot{\rho}} \right)$. 
Identifying $\chi_1$ with the inflaton and $\chi_2$ with the axion, one finds that as long as the axion is frozen in by the Hubble damping, the angle $\theta$ is constantly zero. The inflaton is already the curvature field and the axion the isocurvature one. Further more, with $\theta=0$, the equation of motion for axion perturbations (in Fourier space) simplifies to 
\begin{equation}
  \ddot{\delta\Ax} + 3H\dot{\delta\Ax} + \left(\frac{k^2}{a^2}+m_\Ax^2\right) \delta\Ax = 0.
  \label{eq:eom-axion-pert}
\end{equation}

\section{Squeezed States}
\label{sec:squeezing}
We follow \cite{Ferreira} and especially \cite{Polarski:1995jg} in the application of the notion of squeezed states on primordial cosmic perturbations.
Here we briefly review the formalism based on \cite{Polarski:1995jg} using the example of a generic massless field $\chi$ on an FRW background. It is convenient to use conformal time $\T$ and define the variable $y(\T) \equiv a(\T)\chi(\T)$. The Lagrangian density for $y$ can be written after integration by parts as 
\begin{equation}    
    \L(\T,\vec{x}) = \frac{1}{2} \left( y'^2 - (\Del{i} y)^2 -
    2\frac{a'}{a}yy' + \frac{a'^2}{a^2}y^2 \right), \label{eq:Lagrangian-massless-field}
\end{equation}
where primes denote derivatives with respect to conformal time.
With the momentum $p(\T) \equiv \frac{\partial\tilde{\L}}{\partial y'} = y'-\frac{a'}{a}y$
one can construct the Hamiltonian (in Fourier space) as 
\begin{equation}
  \mathscr{H} = \int \frac{\D^3k}{2} \left( p_\vec{k}p_\vec{k}^* + k^2y_\vec{k}y^*_\vec{k} 
  + \frac{a'}{a} (y_\vec{k}p^*_\vec{k} + p_\vec{k}y^*_\vec{k})\right).
  \label{eq:Hamiltonian-massless-field}
\end{equation}
The time dependence of coordinate and momentum then are
\begin{equation}
  \begin{aligned}
    y' &= \frac{\partial \mathscr{H}}{\partial p} = p + \frac{a'}{a}y \\
    p' &= -\frac{\partial \mathscr{H}}{\partial y} = -ky - \frac{a'}{a}p.
  \end{aligned}
  \label{eq:eom-hamilton-massless-field}
\end{equation}
Now the field is quantised and one defines the common ladder operators:
\begin{equation}
  \begin{aligned}
    \hat{y}_{\vec{k}}(\T) &= \frac{1}{\sqrt{2}} \left( \hat{a}_{\vec{k}}(\T) +
    \hat{a}^\dagger_{-\vec{k}}(T) \right), \\
    \hat{p}_{\vec{k}}(\T) &= -i\sqrt{\frac{k}{k}} \left( \hat{a}_{\vec{k}}(\T) -
    \hat{a}^\dagger_{-\vec{k}}(T) \right).
  \end{aligned}
  \label{eq:y-p-ladder-operators}
\end{equation}
From the canonical commutation relation for $\hat{y}$ and $\hat{p}$, the ladder operators
inherit their usual commutator $[\hat{a}_{\vec{k}}(\T),
\hat{a}^\dagger_{\vec{k}'}] = \delta^{(3)}(\vec{k}-\vec{k}')$. One should note
that the Heisenberg picture is employed here, where the time-dependence is
associated with the operators and the states remain time-independent. 

To find
the time evolution of the ladder operators, Eq.~(\ref{eq:y-p-ladder-operators})
is inserted into (\ref{eq:eom-hamilton-massless-field}). This yields:
\begin{equation}
  \begin{aligned}
    \hat{a}'_{\vec{k}}(\T) &= -ik \hat{a}_{\vec{k}}(\T) + \frac{a'}{a} \hat{a}^\dagger_{-\vec{k}}(\T) \\
    \hat{a}^{\dagger\prime}_{-\vec{k}}(\T) &= \frac{a'}{a} \hat{a}_{\vec{k}}(\T) + ik\hat{a}^\dagger_{-\vec{k}}(\T).
  \end{aligned}
  \label{eq:eom-ladder-operators}
\end{equation}
An ansatz to solve these coupled differential equations is
\begin{equation}
  \begin{aligned}
    \hat{a}_{\vec{k}}(\T) &= \alpha_k(\T)\hat{a}_{\vec{k}}(\T_0) +
    \beta_k(\T)\hat{a}^\dagger_{-\vec{k}}(\T_0),\\
    \hat{a}^\dagger_{-\vec{k}}(\T) &=
    \alpha^*_k(\T)\hat{a}^\dagger_{-\vec{k}}(\T_0) +
    \beta^*_k(\T)\hat{a}_{\vec{k}}(\T_0), 
  \end{aligned}
  \label{eq:parametrisation-ladder-operators}
\end{equation}
which constitutes a Bogolubov transformation of the initial time ladder
operators. As the commutation relations for the ladder operators shall hold for
all times, one finds $|\alpha_k(\T)|^2 - |\beta_k(\T)|^2 =1$. This allows to
parametrise the coefficient functions $\alpha_k(\T)$ and $\beta_k(\T)$ as
\begin{equation}
  \begin{aligned}
    \alpha_k(\T) &= \Exp{-i\theta_k(\T)}\cosh(r_k(\T)), \\
    \beta_k(\T) &= \Exp{i[\theta_k(\T)+2\varphi_k(\T)]}\sinh(r_k(\T)),
  \end{aligned}
  \label{eq:def-squeezing-parameter}
\end{equation}
where $r_k$, $\varphi_k$ and $\theta_k$ are the \emph{squeezing parameters} of interest. 

To find the time evolution of the squeezing parameters, one first inserts Eq.~\eqref{eq:parametrisation-ladder-operators} into Eq.~\eqref{eq:eom-ladder-operators}. This yields
\begin{equation}
  \alpha'_k =\frac{a'}{a}\beta^*_k - ik\alpha_k\, , \quad
  \beta'_k = \frac{a'}{a}\alpha^*_k - ik\beta_k.
  \label{eq:eom-coefficent-functions}
\end{equation}
The derivatives of the coefficient functions are
\begin{equation}
  \begin{aligned}
    \alpha'_k &= \Exp{-i\theta_k} \left[r'_k\sinh(r_k)-i\theta'_k\cosh(r_k)\right]
    \\
    \beta'_k &= \Exp{i(\theta_k(\T)+2\varphi_k(\T))} \left[ r'_k\cosh(r_k) 
    + i(\theta'_k+2\varphi'_k)\sinh(r_k)\right].
  \end{aligned}
  \label{eq:derivatives-coefficent-functions}
\end{equation}
Together these identities can be used in appropriate combinations [namely
$\tfrac{1}{2}(\Exp{i\theta_k}\alpha'_k + \Exp{-i\theta_k}\alpha'^*_k)$, 
$\tfrac{1}{2}(\Exp{i\theta_k}\alpha'_k - \Exp{-i\theta_k}\alpha'^*_k)$
and $\tfrac{1}{2}(\Exp{-i(\theta_k+2\varphi_k)}\beta'_k +
\Exp{i(\theta_k+2\varphi_k)} \beta'^*_k)$] to derive the time dependence of the
squeezing parameters. The result is
\begin{equation}
  \begin{aligned}
    r'_k &= \frac{a'}{a}\cos(2\varphi_k), \\
    \varphi'_k &= -k - \frac{a'}{a}\coth(2r_k)\sin(2\varphi_k), \\
    \theta'_k &= k + \frac{a'}{a}\tanh(r_k)\sin(2\varphi_k).
  \end{aligned}
  \label{eq:eom-squeezing-parameters}
\end{equation}
With the mode function $f_k(\T)=\tfrac{1}{\sqrt{2k}}(\alpha_k(\T)+\beta^*_k(\T))$, one can write the field operator as
\begin{equation}
  \begin{aligned}
    \hat{y}_{\bm{k}}(\T) &= f_k(\T)\hat{a}_{\bm{k}}(\T_0) + f^*_k(\T) \hat{a}^\dagger_{-\bm{k}}(\T_0) \\
    &= \sqrt{2k}\Re(f_k(\T))\hat{y}_{\bm{k}}(\T_0) - \sqrt{\frac{2}{k}}\Im(f_k(\T))\hat{p}_{\bm{k}}(\T_0). \\
  \end{aligned}
  \label{eq:y-squeeze}
\end{equation}

\section{Squeezing the Axion}
\label{sec:results}
\begin{figure*}[t!]
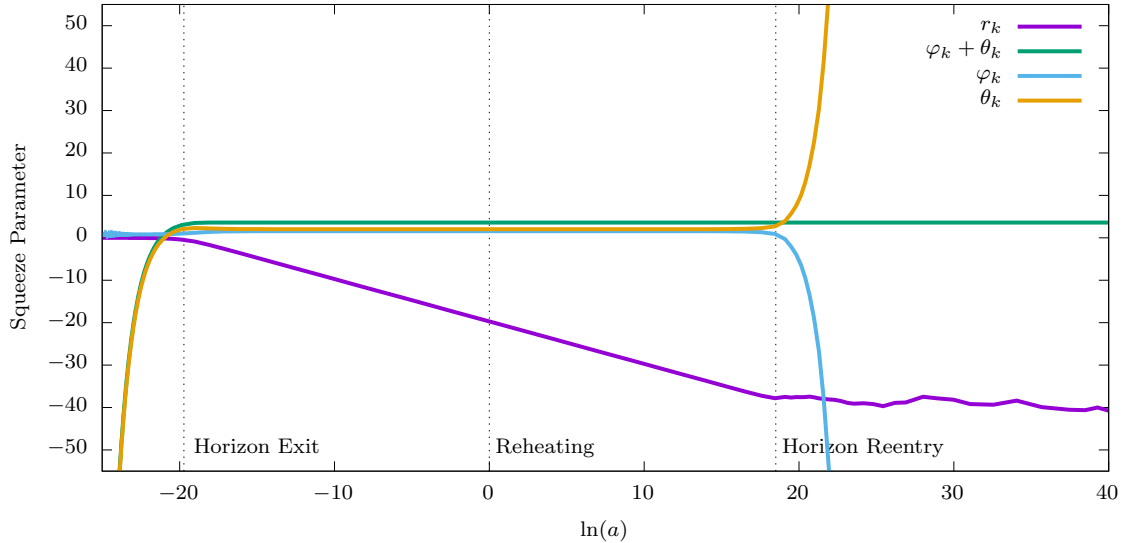

  \centering
  \include{figures/paper-Axion-after-Reheating}
  \caption{
  Evolution of the squeezing parameters of axion
perturbations for a given arbitrarily chosen $k$ mode during inflation and radiation domination
The mode in question leaves the horizon at $\ln(a)\approx -19.7$ and enters again at
$\ln(a)\approx 18.5$. The state squeezes during inflation and continues
becoming more and more squeezed after reheating. After reentering the horizon,
the squeeze factor $r_k$ remains a large constant. The total phase also remains a constant after horizon re-entry. (There are numerical artifacts in the late time evolution of $r_k$, since the rapidly growing $\varphi_k$ enters $r_k'$ via $\cos$. Early time numerical artifacts in the evolution of $\varphi_k$ occur due to limited temporal resolution.)
}
  \label{fig:sq-params-axion-after-reheating}
\end{figure*}
We now apply the formalism of squeezed states to the evolution of axion perturbations during inflation and early radiation domination. To do so, we derive the equations of motions for the squeeze parameter $r_k$, $\varphi_k$ and
$\theta_k$ for the axion and solve them numerically.

The axion's equation of motion, Eq.~\eqref{eq:eom-axion-pert}, corresponds to the action
\begin{equation}
  \begin{aligned}
    S &= \int\D^4x \sqrt{-g} \left( -\frac{1}{2}\Del{\mu}\DEL{\mu}\delta\Ax
    -\frac{1}{2}m_\Ax^2\delta\Ax^2 \right) \\
    &= \int\D\tau\D^3x a^4 \left( \frac{1}{2a^2} \delta\Ax'^2
    -\frac{1}{2a^2}(\nabla\delta\Ax)^2 -\frac{1}{2}m_\Ax^2\delta\Ax^2 \right).
  \end{aligned}
  \label{eq:action-axion-pert}
\end{equation}
Again, it is useful to define $u = a\Ax$.
After integration by parts, the action becomes 
\begin{equation}
  S= \int\D\T\D^3x \frac{1}{2}\left[ u'^2 - (\Del{i} u)^2 
  -\left( m_\Ax^2a^2-\frac{a''}{a} \right) u^2 \right].
  \label{eq:action-a*axion-pert}
\end{equation}
To bring it to a form similar to Eq.~\eqref{eq:Lagrangian-massless-field} one
subtracts the total derivative $\left(\tfrac{a'}{a}u^2\right)'$, yielding
\begin{equation}
\begin{aligned}
  S = &\frac{1}{2} \int\D\T\D^3x\\ &\left[ u'^2 -(\Del{i} u)^2 - \left( m_\Ax^2a^2
  - \left(\frac{a'}{a}\right)^2  \right)u^2 - 2\frac{a'}{a}uu' \right].
  \label{eq:action-a*axion-pert-alt}
\end{aligned}
\end{equation}
The corresponding Hamiltonian then is
\begin{equation}
\begin{aligned}
  \hat{\mathscr{H}} = \int\frac{\D^3k}{2}\Bigg( \hat{p}_{\vec{-k}}\hat{p}_{\vec{k}} &+
  \underbrace{(k^2+m_\Ax^2a^2)}_{m^2_\text{eff}}\hat{u}_{\vec{-k}}\hat{u}_{\vec{k}}\\
  &+  \frac{a'}{a}(\hat{p}_{\vec{-k}}\hat{u}_{\vec{k}}
  + \hat{u}_{\vec{-k}}\hat{p}_{\vec{k}}) \Bigg)
  \label{eq:Hamiltonian-axion-pert}
\end{aligned}
\end{equation}
with $\hat{p}_{\vec{k}} = \hat{u}'_{\vec{k}} - \frac{a'}{a}\hat{u}_{\vec{k}}$.
One finds $\hat{u}' = \hat{p} + \tfrac{a'}{a}\hat{u}$ and
$-\hat{p}' = m^2_{\text{eff}}\hat{u} + \tfrac{a'}{a}\hat{p}$. Introducing ladder operators as in
Eq.~(\ref{eq:y-p-ladder-operators}), their equations of motion are:
\begin{equation}
\begin{aligned}
  \hat{a}'_{\vec{k}}(\T) =
    &-\frac{i}{2} \left(k+\frac{m_\text{eff}^2}{k}\right) \hat{a}_{\vec{k}}(\T) \\ 
   &+ \left[\frac{a'}{a} + \frac{i}{2}\left(k-\frac{m_\text{eff}^2}{k}\right) \right]
      \hat{a}^\dagger_{-\vec{k}}(\T) \\
  \hat{a}^{\dagger\prime}_{-\vec{k}}(\T) = 
    &\left[\frac{a'}{a} - \frac{i}{2}\left(k-\frac{m_\text{eff}^2}{k}\right) \right]
      \hat{a}_{\vec{k}}(\T) \\ 
   &+ \frac{i}{2} \left(k+\frac{m_\text{eff}^2}{k}\right)
      \hat{a}^\dagger_{-\vec{k}}(\T).
  \end{aligned}
  \label{eq:eom-ladder-axion-pert}
\end{equation}
Now, it will be useful to define
\begin{equation}
  \begin{aligned}
    \Omega_k &= \frac{k}{2}+\frac{m^2_{\text{eff}}}{2k} \\
    \lambda_k &= \left( \left( \frac{k}{2}-\frac{m^2_{\text{eff}}}{2k} \right)^2 + 
      \left( \frac{a'}{a} \right)^2 \right)^{\frac{1}{2}} \\
    \phi_k &= -\frac{\pi}{2} + \frac{1}{2}\arctan\left( \frac{a}{a'}
      \left[ \frac{k}{2}-\frac{m^2_{\text{eff}}}{2k} \right) \right]
  \end{aligned}
  \label{eq:params-Ferreira-axion-pert}
\end{equation}
and we rewrite the expression
$\frac{a'}{a} + \frac{i}{2} \left(k-\frac{m_\text{eff}^2}{k}\right)$ as
$-\lambda_{\bm{k}} \exp(i\arctan{x})$ where 
$x= \frac{1}{2}\frac{a}{a'} \left(k-\frac{m_\text{eff}^2}{k}\right)$. Then one can apply a parametrisation as in Eq.~(\ref{eq:parametrisation-ladder-operators}) and
(\ref{eq:def-squeezing-parameter}) to obtain
\begin{equation}
  \begin{aligned}
    r'_k &= -\lambda_k \cos[2(\phi_k-\varphi_k)] \\
    \varphi'_k &= -\Omega_k - \frac{\lambda_k}{2} 
        [\tanh(r_k)+\coth(r_k)] \sin[2(\phi_k-\varphi_k)] \\
      &= -\Omega_k - \lambda_k \coth(2r_k) \sin[2(\phi_k 
        -\varphi_k)] \\
    \theta'_k &= \Omega_k + \lambda_k\tanh(r)\sin[2(\phi_k
      - \varphi_k)].
  \end{aligned}
  \label{eq:eom-sq-params-axion-pert}
\end{equation}
as the equation of motion for the squeezing parameters of the axion
perturbations.

The initial conditions for the squeezing parameter $r_k$, $\varphi_k$ and
$\theta_k$ can be inferred by comparing the mode function of the perturbations,
\begin{equation}
  \begin{aligned}
    f_k(\T) = \frac{1}{\sqrt{2k}} &\big[ \alpha_k(\T)+\beta_k^*(\T) \big] \\
    = \frac{1}{\sqrt{2k}} &\Big[\,\Exp{-i\theta_k(\T)} \cosh(r_k(\T)) \\
    &+ \Exp{-i\left(\theta_k(\T)+2\varphi_k(\T)\right)} \sinh(r_k(\T))\,\Big],
  \end{aligned}
  \label{eq:mode-function-squeezed}
\end{equation}
to the Bunch-Davies vacuum \citep{Baumann:2009ds}
\begin{equation}
  v_k(\T) = \frac{\Exp{-ik\T}}{\sqrt{2k}} \left( 1-\frac{i}{k\T} \right).
  \label{eq:bunch-davies-mode-function-2}
\end{equation}
In slight contrast to \cite{Ferreira}, we find that the mode function coincides to the Bunch-Davies vacuum for:
\begin{equation}
  \begin{aligned}
    r_k &= \arsinh\left(\frac{1}{2k\T}\right)\\
    \varphi_k &= \frac{\pi}{4} - \frac{1}{2}\arctan\left(\frac{1}{2k\T}\right)\\
    \theta_k &= k\T + \arctan\left(\frac{1}{2k\T}\right).
  \end{aligned}
  \label{eq:bunch-davies-squeezed}
\end{equation}

Now we numerically integrate equations (\ref{eq:eom-sq-params-axion-pert}). We assume the axion mass to be  $m_\Ax=1\,$meV, roughly at the upper end of the possible mass range for a QCD axion. The result is presented in Fig.~\ref{fig:sq-params-axion-after-reheating} for a particular mode with $k=10^5\,$GeV.\footnote{This mode re-enters the horizon when $H\approx 10^{-5}\text{ GeV}$ (Fig.~\ref{fig:background}), long before the axion begins to oscillate. The mode that crosses the horizon when the axion begins to oscillate is the smallest $k$ (largest scale) for which our calculation holds. The physical size of such a mode today is $k_m\approx 10^{10}\text{ Mpc}^{-1}(m_\Ax/\text{meV})^{1/2}$, close to the axion Jeans scale \citep[see e.g.][]{Marsh:2015xka,axiverse,Bauer:2020zsj}. The reference scale $k=10^5\text{ GeV}$ corresponds to the Jeans scale for $m_\Ax=10^{-5}\text{ GeV}$. The largest scale for which our results could possibly be applied corresponds to ``fuzzy DM'' with $m_\Ax\approx 10^{-22}\text{ eV}\Rightarrow k_m\approx 10\text{ Mpc}^{-1}$.}One sees that $r_k$ grows (in modulus) when the corresponding mode leaves the horizon, i.e. the state becomes squeezed. Meanwhile $\phi_k$ and $\theta_k$ approach constant values. This pattern does not change with reheating. When the mode reenters the Horizon, $\phi_k$ and $\theta_k$ grow in modulus, but their sum remains constant. Accordingly, $r_k$ stays large, and the state does not become unsqueezed. The growth of the squeeze parameter to large values is consistent with the growth in comoving particle number density while the field remains frozen by Hubble friction. 

That the axion field does not becomes unsqueezed upon horizon reentry in the radiation epoch is 
a result of physical interest. Squeezing initially becomes large when a mode exits the horizon, which happens due to the shrinking of $aH$. However, we observe that squeezing of superhorizon modes continues when the evolution of $aH$ turns around at reheating. Subsequently, squeezing ceases, but does not reverse, when the mode reenters the horizon. The simple statements ``modes squeeze on horizon exit'', or ``modes squeeze as the comoving horizon shrinks'' cannot be naively time reversed. The radiation epoch and horizon reentry are not time reverses of inflation for squeezed states, despite superficial time reversed similarities.

The squeezed axion field can be modelled as a classical stochastic field with fixed phase at late times, and the commutator can be set to zero. This is because for large squeezing (large $r_k$) one can combine the
two angles from Eq.~(\ref{eq:eom-squeezing-parameters}) to get
$\lim_{|r_k|\to\infty}(\theta_k+\varphi_k)'=0$, as both $\coth(2r_k)$ and
$\tanh(r_k)$ approach $1$ in this limit. Hence, one has
$\theta_k+\varphi_k\to\delta_k$ for a constant phase $\delta_k$ (Fig.~\ref{fig:sq-params-axion-after-reheating} shows this clearly). For the
function $f_k$, it then follows \citep{Polarski:1995jg}
\begin{equation}
  f_k(\T) \to \frac{1}{\sqrt{2k}}\Exp{-i\delta_k}\Exp{r_k(\T)}\cos(\varphi_k(\T))
  \label{eq:f-squeezed}
\end{equation}
The constant factor $\Exp{-i\delta_k}$ can be removed by a time-independent
phase rotation, hence $f_k$ can be made real for all
times $\T$ in the limit of large squeezing ($r_k\rightarrow \infty$). Equation (\ref{eq:y-squeeze}) then
shows that all information about $\hat{p}_{\bm{k}}(\T_0)$ vanishes from $\hat{y}_{\bm{k}}(\T)$. Similarly, it can be shown that $\hat{p}_{\bm{k}}(\T) = \sqrt{k/2}\Re{}g_k(\T)\hat{p}_k(\T_0) + \sqrt{2k}\Im{}g_k(\T)\hat{y}_k(\T_0)$ with $g_k = i(f'_k-\tfrac{a'}{a}f_k)$ \citep[cf.][]{Polarski:1995jg}.
Hence, the commutator 
\begin{equation}
    [\hat{y}_{\bm{k}}(\T),\hat{p}_{\bm{k}}(\T)]
    = 2k f_k(f'_k-\tfrac{a'}{a}f_k) [\hat{y}_k(\T_0), \hat{y}_k(\T_0)] = 0
    \label{eqn:vanish_commutator}
\end{equation}
vanishes, and the operators behave like classical field and momentum. We note that Eq.~\eqref{eqn:vanish_commutator} only holds in the limit of $r_k\to\infty$; otherwise, there would be additional terms proportional to $[\hat{y}_k(\T_0), \hat{p}_k(\T_0)]$. \emph{Thus we have demonstrated that the perturbations remain classical upon horizon reentry.}

\begin{figure}[b]
  \centering
  \input{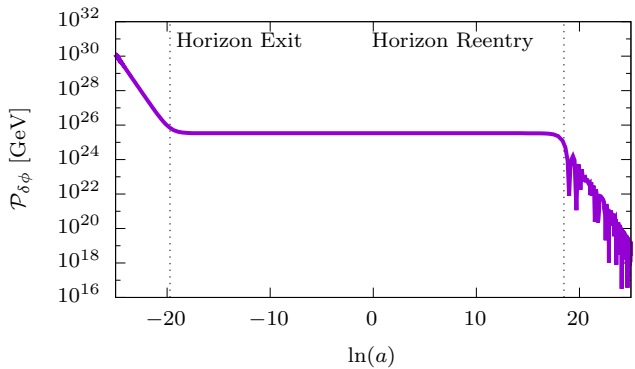}
  \caption{Evolution of the power spectrum of axion perturbations ($k=10^5$\,GeV) during
  inflation and radiation domination. }
  \label{fig:power-spectrum-axion-radiation}
\end{figure}

Lastly we check that our formalism reproduces the expected results of approximate scale invariance for the isocurvature spectrum, and the subsequent classical evolution of the axion field. From the squeezing parameters, we calculated the power spectrum of the axion perturbations by $\P= \frac{k^3}{2\pi^2}\frac{|f_k^2|}{a^2}$. Fig.~\ref{fig:power-spectrum-axion-radiation} shows the time evolution of the mode $k=10^5\,$GeV. One sees that the power spectrum remains constant outside the horizon and decays upon reentry. 

\begin{figure}
  \centering
  \input{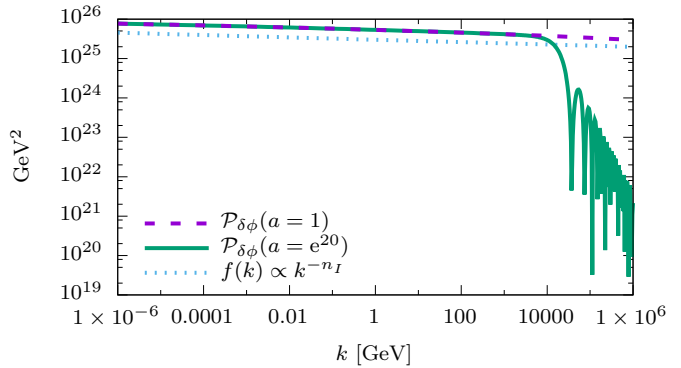}
  \caption{Power spectrum of axion perturbations $\delta\Ax$
    as a function of $k$ at reheating and later for $a=\Exp{20}$. For comparison, a power-law function proportional to $k^{-n_I}$ with $n_I=0.03$ is shown.}
  \label{fig:power-spectrum-axion-k}
\end{figure}

Evaluating the spectrum for different $k$ at reheating ($a=1$) yields approximate scale-invariance, as Fig.~\ref{fig:power-spectrum-axion-k} demonstrates. The spectrum can be fit by the expected slow-roll result: 
\begin{equation}
    P = \frac{H_*^2}{4\pi^2}\left(\frac{k}{k_*}\right)^{-n_I}\, ,
\end{equation}
where we choose the pivot scale $k_*=1$\,GeV for which $H_*\approx4.9\E{13}$\,GeV is the Hubble scale when this mode leaves the horizon (at $a=\Exp{-31.5}$, see Fig.~\ref{fig:background}), and the isocurvature spectral index is $n_I=2\epsilon$ and $\epsilon=-\frac{\dot{H}}{H^2}\approx0.016$ is the slow roll parameter \citep[see e.g.][]{2009ApJS..180..330K}. Hence, we recover the expected behaviour for isocurvature initial power spectrum in slow roll inflation. We also show the spectrum evaluated at a later time, $a=e^{20}$, when small scale modes have re-entered the horizon, showing the expected suppression and oscillatory features below the horizon scale found in the standard classical evolution~\citep{khlopov_scalar,Hlozek:2014lca,Hlozek:2017zzf}.

\section{Conclusion}
\label{sec:Discussion}

We have computed the squeeze parameter of an axion-like scalar DM field during inflation and early radiation domination. The field picks up isocurvature perturbations from quantum vacuum fluctuations during inflation. These perturbations become squeezed when the wavelength becomes larger than the cosmological horizon, and as such the field and momentum operators can be rotated to a basis where the canonical commutators vanish. The scalar field is assumed to be stable and form the cosmic DM. Therefore, perturbations reenter the horizon after reheating. In this regime mode functions begin to oscillate, however the overall phase remains constant, as does the squeeze parameter. Thus, the state of scalar field DM inflationary vacuum fluctuations is well described by a squeezed pure state on all scales that re-enter the horizon while the field is effectively massless, $m_\Ax\ll H$.

We reiterate that the modes we have studied are expected to be the ``most quantum'' in axion cosmology. These modes are generated from vacuum fluctuations, and have not interacted with the thermal bath in any way. These modes re-enter the horizon before the scalar field mass becomes relevant, and  correspond to scales below the cosmic de Broglie wavelength (Jeans scale). 

Our calculation is valid as long as the background field is frozen by Hubble friction, that is, as long as the scalar field/axion mass is small compared to the Hubble parameter, $m_\Ax\ll H$. In our model, this happens around $\ln(a)=28$; the largest modes that reenter the horizon before this are at the order of $10$\,GeV, which redshifted to today correspond to $\approx1\E{10}\,\text{Mpc}^{-1}$ 
When $\Ax$ unfreezes and oscillates, interactions between the adiabatic and the isocurvature mode give rise to mixing between the axion perturbations and the thermal bath on all scales, mediated by gravity. In particular, this mixes isocurvature and curvature~\citep{Gordon:2000hv} converting one into the other on large scales and at late times. Even in linear cosmological perturbation theory, this interaction mixes the axion field with the radiation bath \citep[thus generating a non-zero curvature from the initial isocurvature state, see e.g. power series of isocurvature initial states in][]{Bucher:1999re,Hlozek:2017zzf}, and might be expected to give rise to decoherence of the isocurvature mode even before the non-linear regime~\citep[the same mechanism occurs in the curvaton model,][except that unlike the curvaton the axion does not decay]{Lyth:2001nq}. 

Scalar field DM fluctuations in the orthogonal adiabatic mode are generated solely from gravitational interactions with the radiation bath, and so must be described by a classical stochastic model. Furthermore, super-Jeans modes eventually grow and undergo non-linear gravitational collapse into galactic DM halos. \cite{Allali:2020ttz} has shown that gravitational scattering of DM with hydrogen inside galaxies indeed leads to decoherence. If the field has non-gravitational interactions these also becomes relevant when $m_\Ax\ll H$ (when the shift symmetry is broken), and will lead to additional decoherence.

One possible application of our result is to provide the initial correlation state of the axion field in the pre-inflation symmetry breaking scenario. Another application of our result could be to provide the correct initial quantum state for the Peccei-Quinn field in the scenario when the $U(1)_{\rm PQ}$ symmetry is broken after inflation, simply by replacing what we have termed the axion with two massless real fields. In this case, the subsequent evolution is highly non-linear, although governed by local interactions in the field potential rather than by gravity \citep[see e.g.][]{Berges:2014xea,Gorghetto:2018myk,Vaquero:2018tib}.

The present work can be seen as a quantum mechanical interpretation of well-known results in linear perturbation theory of scalar field/axion DM. It would be fruitful to further consider the quantum mechanical interpretation of non-linear phenomena, such as the axion bispectrum \citep[e.g.][]{Langlois:2012tm}). Signals of quantum mechanics in cosmology are generically thought to arise as poles in higher-point correlation functions~\citep{Green:2020whw}. However, these ``cosmological collider''~\citep{Arkani-Hamed:2015bza} phenomena are generally present for particles with $m\gg H_I$, while the case we have studied applies in the opposite limit. While the present work was in preparation, \cite{Lu:2021gso} appeared considering such an ``axion isocurvature collider''. It would be interesting to develop the open-EFT/Linblad formalism of Ref.~\cite{Burgess:2014eoa} applied to DM, which extends rigorously the intuition about classicalization derived from the squeezed state formalism.

\emph{Note Added}: In the final stages of preparation, \cite{Kopp:2021ltb} appeared also discussing the squeezing of axion perturbations.

\acknowledgements{We acknowledge useful discussions with Pedro Ferreira, Mark Hertzberg, Erik Lentz, and Eugene Lim. DJEM is supported by an Ernest Rutherford Fellowship from the UK STFC.}

\bibliography{references}

\end{document}